\documentclass[fleqn,twoside]{article}
\usepackage{espcrc2}

\usepackage{graphicx}
\usepackage[figuresright]{rotating}

\newcommand{\AmS}{{\protect\the\textfont2
  A\kern-.1667em\lower.5ex\hbox{M}\kern-.125emS}}

\title{A reinterpretation of Volcano Ranch lateral distribution measurements to infer the mass composition of cosmic rays}

\author{
M. T. Dova\address[UNLP]{Department of Physics, Universidad Nacional de La Plata,
	   La Plata, 1900, Argentina}\thanks{dova@fisica.unlp.edu.ar},
M. E. Mance\~nido\addressmark[UNLP]\thanks{mance@fisica.unlp.edu.ar}, 
A. G. Mariazzi\addressmark[UNLP]\thanks{mariazzi@fisica.unlp.edu.ar},
T. P. McCauley\address{Department of Physics, Northeastern University, Boston, MA 02115, USA}\thanks{mccauley.t@neu.edu}
        and
A. A. Watson\address{Department of Physics and Astronomy, University of Leeds, Leeds, LS2 9JT, UK}\thanks{a.a.watson@leeds.ac.uk}
}
       
\begin{document}

\begin{abstract}
In the course of its operation, the Volcano Ranch array collected
data on the lateral distribution of showers produced by cosmic
rays at energies above $10^{17}$ {\rm eV}. From these data very precise 
measurements of the steepness of the lateral distribution function,
characterized by the $\eta$ parameter, were made. The current 
availability of sophisticated hadronic interaction models has
prompted a reinterpretation of the measurements. We use the
interaction models {\sc qgsjet} and {\sc sibyll} in the {\sc aires} Monte Carlo code to generate showers
together with {\sc geant4} to simulate the response of the detectors to 
ground particles. As part of an effort to estimate the primary mass composition of cosmic rays 
at this energy range, we present the results of our preliminary analysis of the distribution of $\eta$. 
\vspace{1pc}
\end{abstract}

\maketitle

\section{Introduction}
Over its lifetime, the Volcano Ranch array~\cite{vrall} collected data on the lateral distribution of air showers produced by cosmic rays of energies above $10^{17}$ {\rm eV}. This lateral distribution
is parameterized by a so-called lateral distribution function. The steepness of this function, given by 
$\eta$ , is sensitive to the depth of maximum ($X_{\rm max}$) of the shower, and therefore to the primary
composition and to the character of the initial hadronic interactions. 

Our analysis of precise Volcano Ranch measurements of $\eta$ is the first analysis of Volcano Ranch data
using modern Monte Carlo tools. To simulate the development of the air shower we use the 
{\sc aires}~\cite{aires} code (version 2.4.0), which contains the hadronic interaction generators {\sc qgsjet98}~\cite{qgsjet} and 
{\sc sibyll2.1}~\cite{sibyll}. The results of these simulations are convolved with a simulation of
the detector response carried out using {\sc geant4}~\cite{geant4}.

\section{Volcano Ranch}
The pioneering Volcano Ranch array was an array of scintillation counters in operation
from 1959-1974 at the MIT Volcano Ranch station located near Albuquerque, New Mexico. One of its
many distinctions was the detection of the first cosmic ray event with an energy estimated at 
$10^{20}$ {\rm eV}~\cite{vrevent}. Over its lifetime, the array existed in three distinct configurations.
The first configuration consisted of twenty scintillator detectors of surface area 3.26 ${\rm m^{2}}$. Nineteen
detectors were spaced 442 {\rm m} apart. A twentieth detector was placed at various locations
and used for the measurement of the density of particles. A second configuration had a larger spacing of 884 ${\rm m}$. 

It is the third configuration which is most significant for our analysis. In this configuration, the number 
of detectors was quadrupled by splitting up each of the twenty detectors into eighty detectors of 
surface area 0.815 ${\rm m^{2}}$ spaced 147 {\rm m} apart. This configuration, with many more detectors spaced closer
together, allowed for very precise measurements of the lateral distribution of signals in the detectors. The
steepness of the lateral distribution, and fluctuations in it, may lead to an estimate of the primary
composition.

\section{Simulation of detector response}
To simulate the detector response of the array to the ground particles, we utilized the 
general-purpose simulation toolkit {\sc geant4}. Our procedure follows the 
prescription found in ~\cite{kutter}, where the detector response to electrons, gamma, and
muons is simulated in the energy range 0.1 to 1.0e5 {\rm MeV} and for five bins per decade of energy.
This detector response is convolved with the results of {\sc aires} air shower simulations to
obtain scintillator yield in minimum ionizing particles per square meter ${\rm (mips/m^{2})}$.


\section{Lateral distribution function}

For Volcano Ranch data, a generalized version of the Nishimura-Kamata-Greisen (NKG) formula
was used by Linsley to describe the lateral distribution of particles at ground in ${\rm mips/m^{2}}$~\cite{denver}. This lateral distribution function is given as 

\begin{equation}
S_{vr}(r) = {\frac{N}{r_{m}^{2}}} C(\alpha,\eta) \left( {\frac{r}{r_{m}}} \right)^{-\alpha} \left( 1 + {\frac{r}{r_{m}}} \right)^{-(\eta-\alpha)} 
\end{equation}

normalized to $N$ with

\begin{equation}
C = {\frac{\Gamma(\eta-\alpha)}{2\pi\Gamma(2-\alpha)\Gamma(\eta-2)}} .
\end{equation}

Here $r_{m}$ is the Moliere radius, which is $\simeq$ 100 {\rm m} at the Volcano Ranch elevation.

From a subset of showers detected by the array in the third configuration, the following form
of $\eta$ as a function of zenith angle and shower size was found to fit the data~\cite{plovdiv1}: 

\begin{equation}
\langle \eta(\theta, N) \rangle = b_{0} + b_{1}(\sec \theta - 1) + b_{2}\:log_{10}({\frac{N}{10^{8}}})
\end{equation}

with $b_{0} = 3.88 \pm 0.054$, $b_{1} = - 0.64 \pm 0.07$, and $b_{2} = 0.07 \pm 0.03$.

\section{Preliminary comparison of lateral distribution}
\begin{figure}[htb]
\vspace{9pt}
\includegraphics[width=18.5pc]{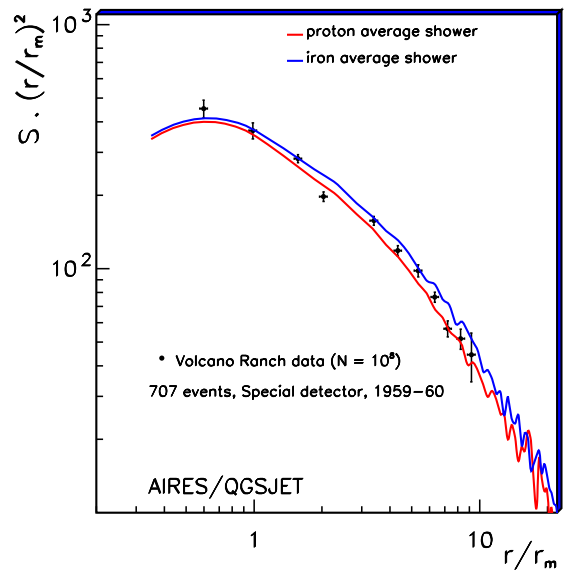}
\caption{Average lateral distribution of simulated showers compared to average measurements from
Volcano Ranch.}
\label{fig:firstLD}
\end{figure}

Here we compare the Volcano Ranch lateral distribution with modern Monte Carlo simulations. 
In Figure~\ref{fig:firstLD} we show how well the simulations ({\sc aires/qgsjet98} with the detector
response included) reproduce the average lateral distribution of 707 showers of estimated size $N = 10^{8}$, measured by the array in its first configuration~\cite{denver}.   
 
\begin{figure}[htb]
\vspace{9pt}
\includegraphics[width=18.5pc]{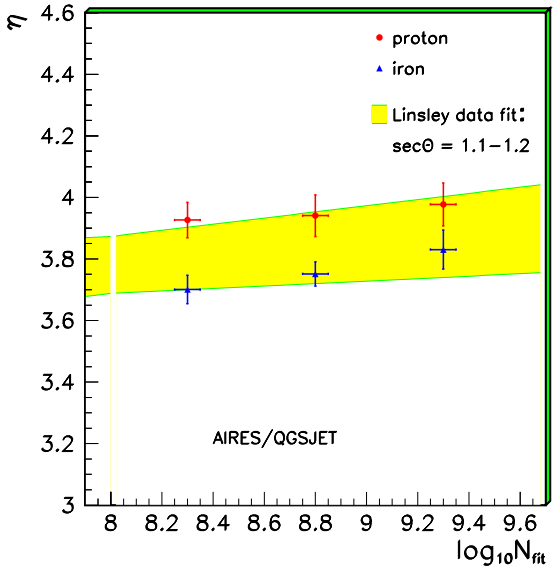}
\caption{Comparison of $\eta$ as a function of shower size for $\sec \theta = 1.1 - 1.2$
and {\sc aires/qgsjet98}}
\label{fig:etaN_qgsjet}
\end{figure}

\begin{figure}[htb]
\vspace{9pt}
\includegraphics[width=18.5pc]{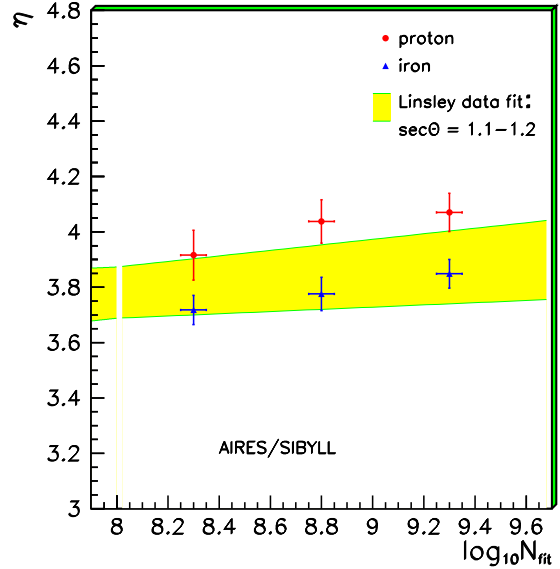}
\caption{Comparison of $\eta$ as a function of shower size for $\sec \theta = 1.1 - 1.2$
and {\sc aires/sibyll2.1}}
\label{fig:etaN_sibyll}
\end{figure}

As a further comparison, for fixed
bins of zenith angle, we determine the number of particles at ground level from a fit to the lateral 
distribution function (with $\alpha = 1$) and compare to the average functional form of
$\eta$ given by Equation 3. The results
of this comparison for $\sec \theta = 1.1 - 1.2$ can be seen for proton and iron showers using 
{\sc aires/qgsjet98} and {\sc aires/sibyll2.1} in Figures~\ref{fig:etaN_qgsjet} 
and~\ref{fig:etaN_sibyll}. 
One can see that the average form of $\eta$ over a realistic range of mass composition, from proton to iron, is reproduced by the 
simulations. At this stage, any attempt to draw a firm conclusion about composition or the suitability of interaction models is premature; we have limited Monte Carlo
statistics and are only comparing to an average functional form of $\eta$. What we are content to take from
this stage of the analysis is encouragement from the fact that the simulation can reproduce the data. A more detailed analysis will follow in a later work.

\section{Towards an estimate of primary composition}
A more robust estimator of primary composition comes from an analysis of the measured 
fluctuations of $\eta$, which were found to be significantly greater than that expected from
measurement error alone~\cite{plovdiv1}.
From our knowledge of the characteristics of showers produced by nucleons and nuclei, one would 
expect that measured fluctuations in the distribution of $\eta$ would be smaller for a 
heavier composition. This can be understood if we consider an air shower produced by a nucleus 
with $A$ nucleons as a superposition of $A$ showers each with energy $E/A$. Thus in iron-initiated
showers the average fluctuations in shower development are reduced.

\begin{figure}[htb]
\vspace{9pt}
\includegraphics[width=18pc]{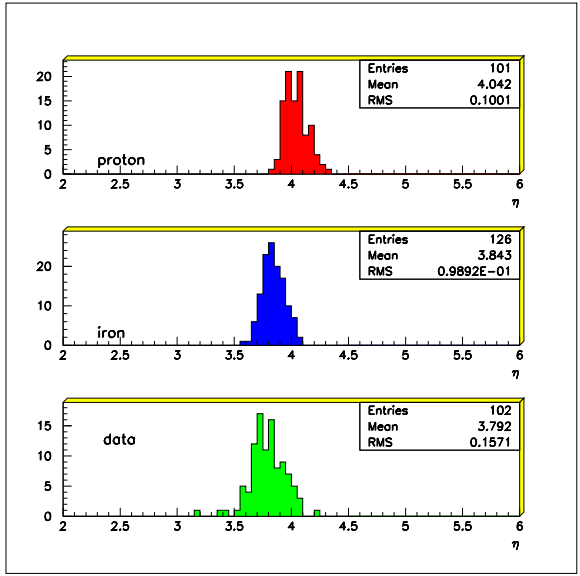}
\caption{Fitted $\eta$ parameter for proton and iron compared to Volcano Ranch data 
with $\alpha=1$, for $\sec \theta$ = 1.0 - 1.1,
using  {\sc qgsjet98}}
\label{fig:1stbin_qgsjet}
\end{figure}

\begin{figure}[htb]
\vspace{9pt}
\includegraphics[width=18pc]{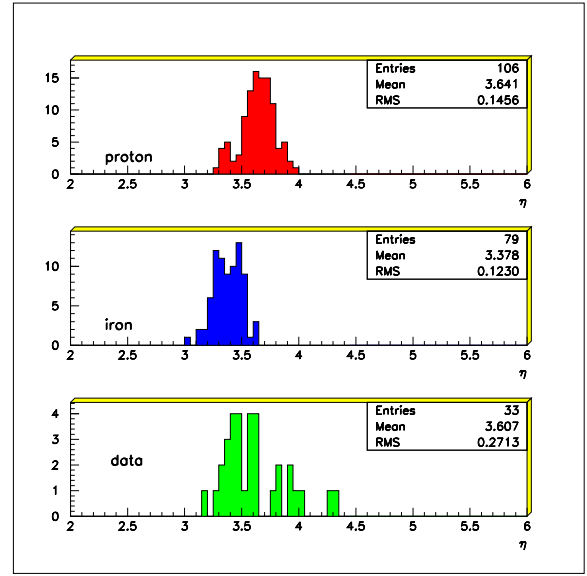}
\caption{Fitted $\eta$ parameter for proton and iron compared to Volcano Ranch data 
with $\alpha=1$, for $\sec \theta$ = 1.4 - 1.5,
using  {\sc qgsjet98}}
\label{fig:2ndbin_qgsjet}
\end{figure}

At Volcano Ranch, measurements of $\eta$ were made on a shower-to-shower basis for fixed
bins of zenith angle~\cite{plovdiv2}. Our procedure is to simulate showers in these bins of zenith angle, 
find $\eta$ from a fit to the lateral distribution function, and compare to data. The results of this
preliminary comparison can be seen in Figures~\ref{fig:1stbin_qgsjet} and~\ref{fig:2ndbin_qgsjet}. 

\section{Conclusions and future plans}
Using modern Monte Carlo tools previously unavailable during its operation
we have been able to reproduce lateral distribution measurements taken at the 
Volcano Ranch array. With an eye towards estimating the primary composition from
these measurements, our ability to reproduce them gives us some confidence that our
analysis procedure is correct and indicates that current hadronic interaction models 
can describe the data fairly well. A preliminary analysis of the distribution of the
steepness of the lateral distribution function, $\eta$, indicates that indeed something can be said about primary composition from an analysis of the data, but that further analysis is needed to
make any stronger statement. However, it is encouraging that the data from Volcano Ranch may be used to say something
useful about mass composition thirty years after its closure.

\end{document}